\documentclass[epj]{svjour}
\usepackage{graphics,amsmath}

\usepackage{graphics,epsfig}
\begin{document}
\title{Deviations from Perfect Memory in Spin Glass Temperature Cycling Experiments}
\subtitle{}

\author{M.~Sasaki\inst{1}, V.~Dupuis\inst{2}, J.-P.~Bouchaud\inst{2} \and E.~Vincent\inst{2}}
\authorrunning{M.~Sasaki {\it et al.}}
\offprints{sasaki@ipno.in2p3.fr}
\institute{Laboratoire de Physique Th\'eorique et Mod\`eles Statistiques,
b\^at. 100, Universit\'e Paris-Sud, F--91405 Orsay, France
\and Service de Physique de l'\'Etat Condens\'e,
Orme des Merisiers --- CEA Saclay, 91191 Gif sur Yvette Cedex, France.}

\date{\today}

\abstract{We study the deviations from perfect memory in negative temperature 
cycle spin glass experiments. It is known that the a.c. susceptibility 
after the temperature is raised back to its initial value is superimposed 
to the reference isothermal curve for large enough temperature
jumps $\Delta T$ (perfect memory). For smaller $\Delta T$, the deviation
from this perfect memory has a striking non monotonous behavior: the `memory
anomaly' is {\it negative} for small $\Delta T$'s, becomes positive for 
intermediate $\Delta T$'s, before vanishing for still larger $\Delta T$'s.
We show that this interesting behavior can be reproduced by simple Random 
Energy trap models. We discuss an alternative interpretation in terms of droplets
and temperature chaos.
}

\maketitle
%
%%%%%%%%%%%%%%%%%%%%%%%%%%%%%%%%%%%%%%%%%%%%%

\section{Introduction}
 
It is well known that in glassy systems, dynamical effects strongly depend on 
the history of the system after quenching from above the glass transition temperature 
$T_g$. These phenomena are called aging and have been studied using 
various experimental protocols\cite{Sitges,books2,books3}. 
The measurement of the ac-susceptibility during negative $T$-cycles is one of them. 
This experiment consists of the following three stages. 
In the first stage, the system 
is quenched from above its critical temperature \( T_g \) and it is kept at 
a temperature $T_1$ ($<T_g$) during a time \( t_{{\rm w}1}\). 
In the second stage the temperature is temporally reduced 
to \(T_2= T_1- \Delta T\) during a time \( t_{{\rm w}2}\), 
and then it is set back to \( T_1 \) in the third stage. 
The ac-susceptibility (magnetic, dielectric, mechanical,..) 
is measured during all the three stages. 
The effect of the perturbation of the temperature is examined 
by comparing the perturbed and unperturbed (i.e. \( t_{{\rm w}2} =0 \)) data 
in the third stage. From this comparison, it is revealed that 
the perturbed data quickly approaches the unperturbed one as if 
the system remembers how far the relaxation at the temperature $T_1$
had proceeded before the perturbation, even though the system is strongly 
{\it rejuvenated} at temperature $T_2$ (see Fig. \ref{Fig:1}). This phenomenon is called the memory 
effect, and has been observed in many glassy systems 
like spin glasses\cite{SGactc1,SGactc2,SGactc3,SGactc4}, 
orientational glasses\cite{OGactc1,OGactc2,OGactc3}, 
polymer glasses\cite{PGactc}, etc. 

In the present paper, we focus on the deviations from perfect memory
that are observed immediately after heating back the system to $T_1$.
Surprisingly, systematic experiments (that we report in section 
\ref{sec:Experiments}) show that the transient behavior is 
non monotonous as a function of $\Delta T$. The initial extra contribution, that
we will call the memory anomaly, 
is found to be {\it negative} for small $\Delta T$, then {\it positive}
for intermediate $\Delta T$, before vanishing completely for large 
enough $\Delta T$ (perfect memory).

We then measure the ac-susceptibility for a $T$-cycle in the 
Random Energy Model (REM)\cite{Derrida1} or the Generalized 
REM (GREM)\cite{Derrida2}. These models 
have been shown\cite{GREM1,GREM2,GREM3,GREM4,GREM5,GREM6} 
to reproduce many of the experimental features of aging,
including rejuvenation and memory. We find that the non monotonous transients
mentioned above can also be obtained in such models. 

Technically, we first
establish a relation between the ac-susceptibility and the distribution of relaxation times. 
Although this relation could not be derived analytically, 
numerical tests suggest that this relation 
holds with high accuracy. This enables us to 
measure the ac-susceptibility for any desired time scale
in the REM (but not in the GREM), allowing us to measure 
the ac-susceptibility in the REM up to time scales comparable to experiments. 
This would not be possible using Monte Carlo simulations. 
As for the GREM, the ac-susceptibility is obtained using Monte Carlo simulation,
and therefore corresponds to rather small time scales. 
Our numerical measurements are made for various sets of parameters, i.e.,  
the waiting times $t_{{\rm w}1}$ and $t_{{\rm w}2}$, the temperatures $T_1, T_2$ and 
the period of the 
applied ac-field $P=2\pi/\omega$. We find that the non monotonous effect
described above depends quite sensitively on some of these parameters.

The organization of this manuscript is as follows: 
In section \ref{sec:Experiments} results of new complete set of $T$-cycle experiments 
in spin glasses are shown. 
In section \ref{sec:Model} we explain the REM, the GREM 
and the dynamics employed for these models. 
In section \ref{sec:measurement} a relation between ac-susceptibility and 
distribution of relaxation times is proposed and its validity is tested 
numerically. In section \ref{sec:resultREM} and \ref{sec:resultGREM}
results on the REM and on the GREM are 
shown. Finally, in section \ref{sec:discussion}, we give a physical
discussion of our results and a comparison with the scenario of temperature chaos,
where the existence of an overlap length is assumed. 

\section{Results of Experiments}\label{sec:Experiments}

\begin{figure}
\centerline{\hbox{\epsfig{figure=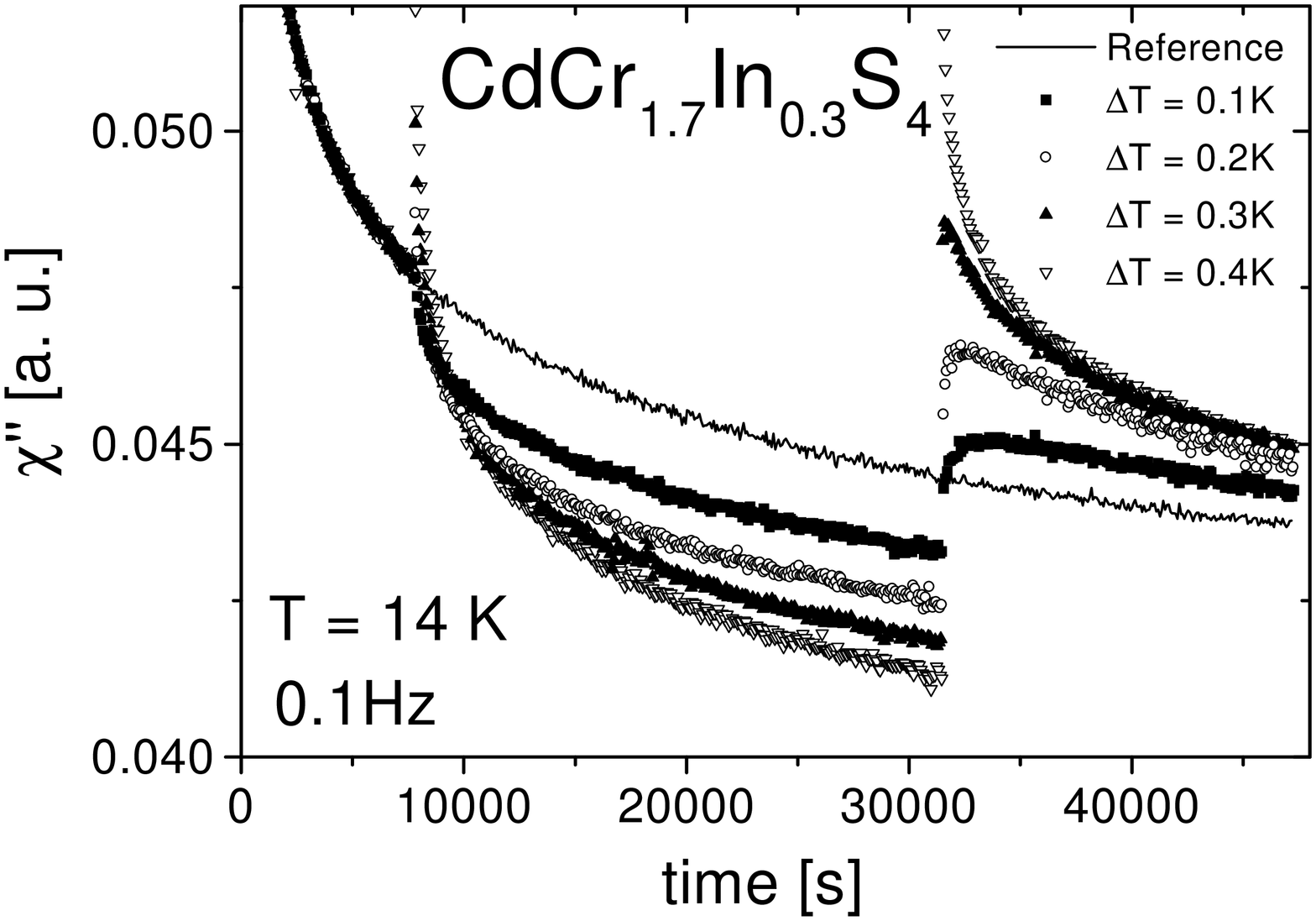,height=8.4cm,angle=270}}}
\caption{Out-of-phase susceptibility $\chi''$ vs. time for temperature cycling
experiments at $T_1=14K$ on the $CdCr_{1.7}In_{0.3}S_4$ Heisenberg spin glass ($T_g=16.7K$). The sample was
quenched to $T_1$, kept at this temperature during $t_1=7700s$, and then submitted to a
negative temperature cycle at $T_1-\Delta T$ during $t_2=23650s$ before returning to $T_1$ for a time
$t_3$. Four different small $\Delta T$ ($0.1K$-$0.4K$) were used.}
\label{Fig:1}
\end{figure}

The effect of temperature changes on aging has already been largely investigated experimentally in
spin glasses \cite{Sitges}. Here we focus on the details of the experimental results 
in the well
characterized thiospinel $CdCr_{1.7}In_{0.3}S_4$ ($T_g=16.7K$) 
Heisenberg spin glass sample.
Fig. \ref{Fig:1} presents the results on the out-of-phase component of $\chi''$ of 
the a.c. susceptibility 
during the temperature cycle described in the introduction, 
with $T_1$=14K, for four $\Delta T$ values in the range ($0.1K$-$0.4K$), 
and a frequency $0.1$ Hz. Note that the period of the
a.c. field ($10$ sec) is typically much larger than any microscopic 
time scale ($10^{-12}$ sec).
\footnote{The relevant `microscopic' time scale may however be strongly 
renormalized by critical
fluctuations, as recently discussed 
in refs.~\cite{SGactc4,PRB,Yoshino:Suedes}, 
and can be much larger than $10^{-12}$ sec. for
$T$ close to $T_g$.}

\begin{figure}
\centerline{\hbox{\epsfig{figure=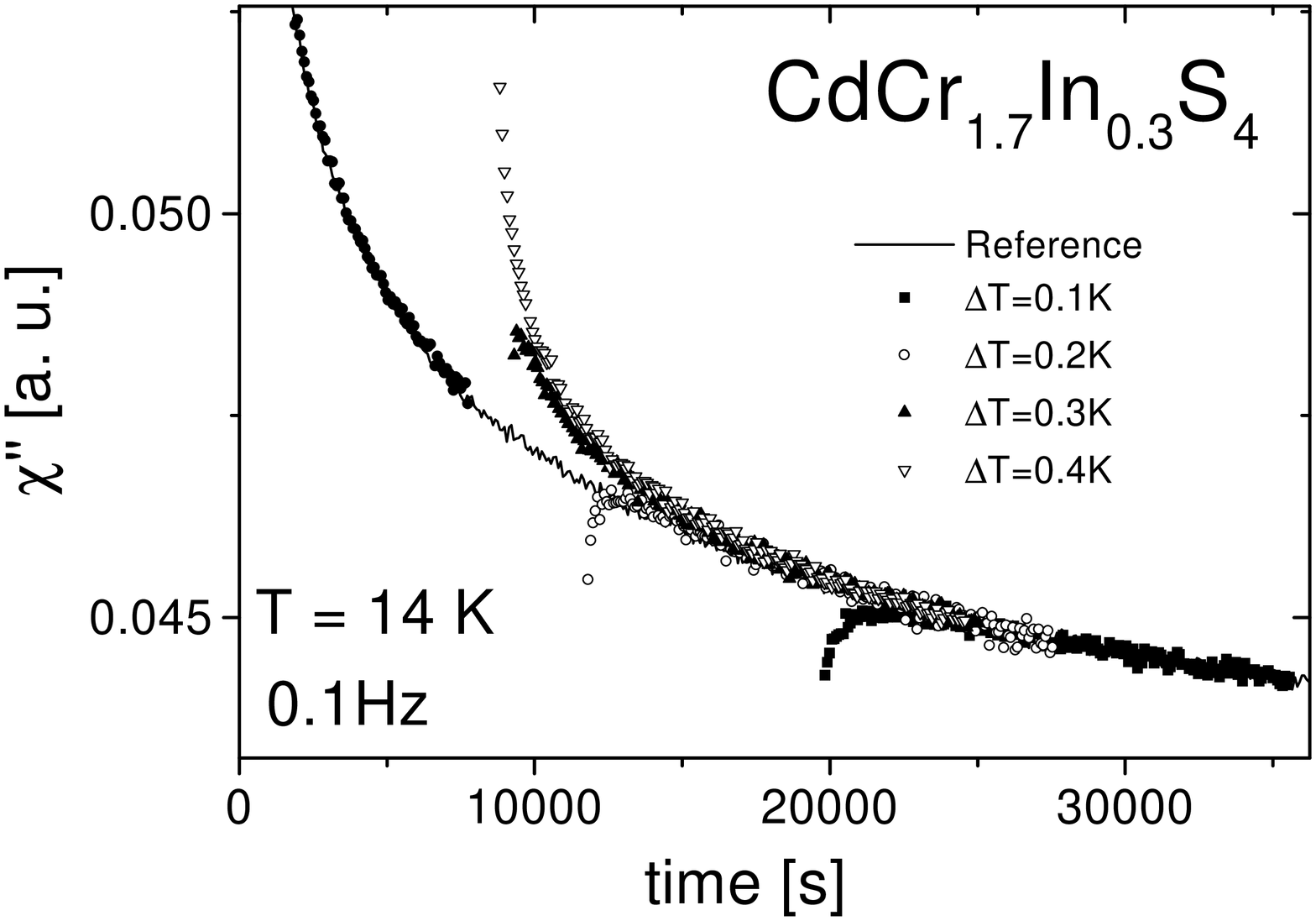,height=8.4cm,angle=270}}}
\caption{Superposition of $\chi''$-relaxations of Fig.1 after the negative temperature
cycles onto a reference isothermal relaxation curve (solid line). The data measured after the cycle have been shifted
horizontally by $t_2-t_{\rm eff}$ to take into account the effective contribution $t_{\rm eff}$ of aging at $T_1-\Delta T$ on aging at
$T_1$. The merging of the data points with the reference curve occurs first from below for small
$\Delta T$'s and then from above for larger $\Delta T$'s.}
\label{Fig:2}
\end{figure}

Just after the initial quench, $\chi''$ is slowly relaxing downward with time 
$t_{{\rm w}1}$ due to aging. When $T_1$ is
decreased to $T_2=T_1-\Delta T$, we observe a jump and a strong relaxation 
in $\chi''$: this is the rejuvenation effect. 
Despite this strong reinitialization of aging at $T_1-\Delta T$, 
it is possible, when
the temperature is raised back to $T_1$ and for large enough $\Delta T$ 
($\ge 2K$) \cite{SGactc1}, 
to find a perfect memory of the past relaxation at $T_1$. 
For large $\Delta T$'s, the relaxations at $T_1$ before and after the
temperature cycle are in exact continuity and there is no contribution 
of aging at $T_1-\Delta T$ on
aging at $T_1$. In contrast, in the regime of small $\Delta T$'s of 
Fig. \ref{Fig:1}, we do not find a perfect
memory of aging after the temperature cycle. The $\chi''$ relaxations after 
the negative cycle can still
be superposed, apart from a transient contribution to be analyzed below, 
onto a reference isothermal relaxation at $T_1$ but
we now need to shift the data by an effective time 
$t_{\rm eff} < t_{{\rm w}2}$ which accounts for aging during the stay at 
$T_2$  (see Fig. \ref{Fig:2}). This effective time $t_{\rm eff}$ has 
been recently studied in detail both experimentally\cite{SGactc4,Yoshino:Suedes} 
and numerically\cite{Yoshino:Takayama,BerthierJPB}.

\begin{figure}
\centerline{\hbox{\epsfig{figure=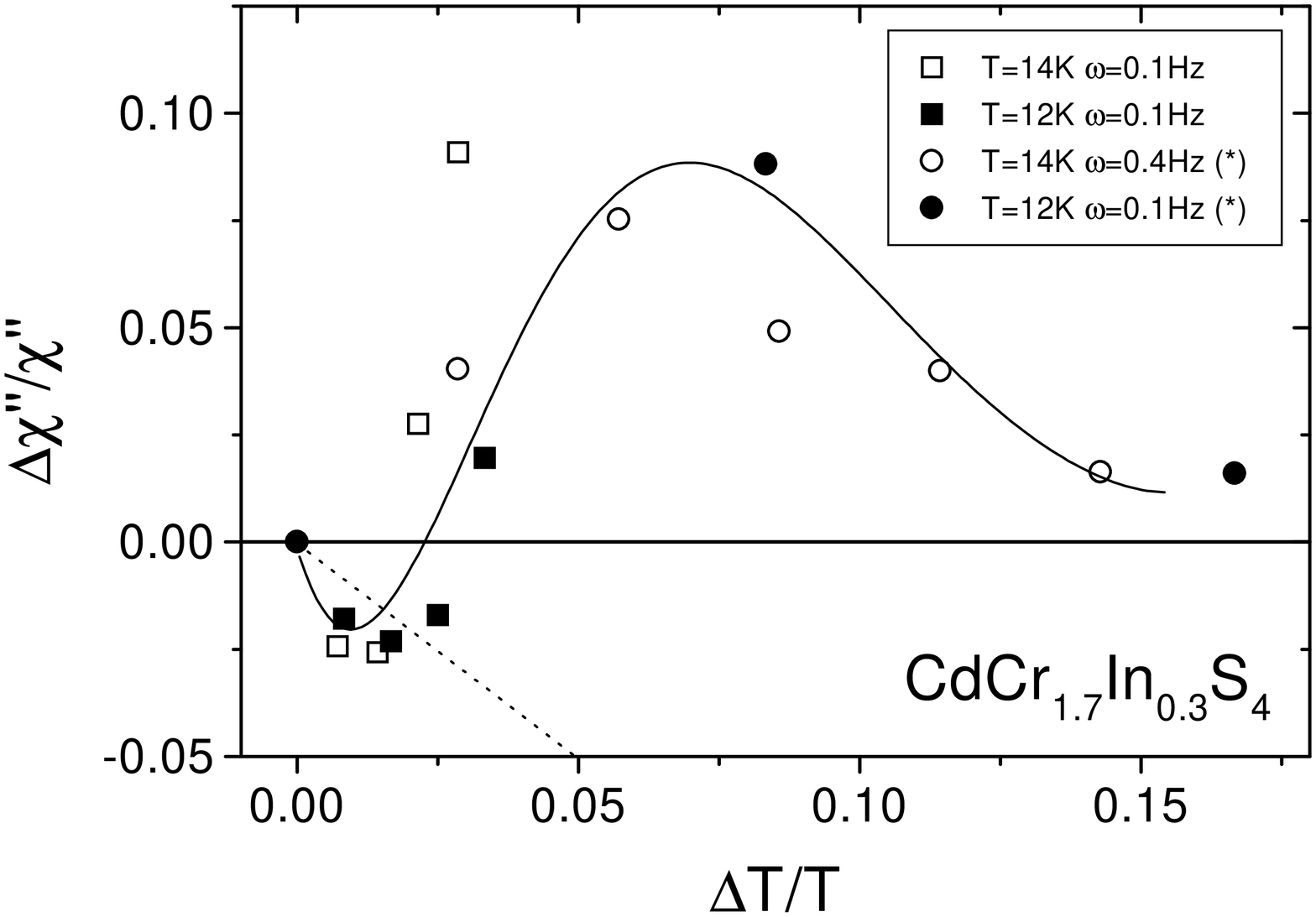,height=8.4cm,angle=270}}}
\caption{Relative difference (memory `anomaly') $\Delta \chi''/\chi''$ between the $\chi''$-value after a
negative temperature cycle of amplitude $\Delta T$ at a temperature $T$ and the $\chi''$-value
after a corresponding isothermal aging at $T$ during a time $t_1+t_{\rm eff}(\Delta T)$ for various
temperature cycling experiments (such as the ones presented in Fig. 1-2). The first
cycles at $14K$ and $12K$ ($0.1Hz$) correspond to $t_1=7700s$ and $t_2=23650$. The others (*)
correspond to $t_1=t_2=3600s$ for $14K$ ($0.4Hz$) and $t_1=t_2=18000s$ for $12K$ ($0.1Hz$). The
thin line is a guide for the eye. The dotted line corresponds to $\Delta \chi''/\chi''=-\Delta T/T$.}
\label{Fig:3}
\end{figure}

Coming back to the transient contribution, we see that for the
smallest $\Delta T=0.1K$ and $0.2K$ used, $\chi''$ reaches a
maximum as a function of time, and merges back with the reference 
curve from below, 
while for $\Delta T=0.3K$ and $0.4K$ the maximum disappears and this return occurs from above. This is a systematic effect
which is also observed at other temperatures. We have further characterized 
this feature by gathering
the results of several negative temperature cycling experiments done on the 
thiospinel sample at
two temperatures $T_1=12K$ and $T_1=14K$ and for various $\Delta T$'s. 
In Fig. \ref{Fig:3}, we have plotted the
(relative) memory `anomaly' $\Delta \chi''/\chi''$, where $\Delta \chi''$ 
is the difference between the $\chi''$-value just after the cycle and the
one corresponding to an isothermal aging at $T$ during $t_{{\rm w}1}+t_{\rm eff}$, 
as a function of $\Delta T/T_1$ for the whole set of available experimental results. The characteristic time scales are given in Tables 1 and 2.

\begin{table}
\caption[]{\small $t_{\rm{eff}}$: shift time,  $t^*$: position of the maximum of 
$\chi''$, $t_{\rm rec.}$: time at which the signal merges with the shifted 
reference curve, and $\Delta \chi''/\chi''$: amplitude of the 
memory anomaly for different $\Delta T$'s, for $T_1=14$ K. 
 }
\begin{center}
\begin{tabular}{||c|c|c|c|c||} \hline\hline
$\Delta T$ (K) &  $t_{\rm{eff}}$ (sec) & $t^*$ (sec) & $t_{\rm rec.}$ (sec) &
$\Delta \chi''/\chi''$ \\ \hline
0.1 &  12000 &  1800  &  4200 &  $-2.4\times 10^{-2}$\\ \hline
0.2  &  4000 &  900  &  900 &  $-2.6\times 10^{-2}$\\ \hline
0.3  &  1500 &   160   &  8000 &  $2.8\times 10^{-2}$\\ \hline
0.4  &  1000 &   --    &  10350 &  $9.1\times 10^{-2}$\\ \hline
\end{tabular}
\label{tab:T14}
\end{center} 
\end{table}

\begin{table}
\caption[]{\small Same caption as Table~1 for $T_1=12$ K.}
\begin{center}
\begin{tabular}{||c|c|c|c|c||} \hline\hline
$\Delta T$ (K) & $t_{\rm{eff}}$ (sec) & $t^*$ (sec) &
$t_{\rm rec.}$  (sec)
&
$\Delta \chi''/\chi''$ \\ \hline
0.1 &  15000 &  1200  &  2850 &  $-1.8\times 10^{-2}$\\ \hline
0.2  &  8000 &  1200  &  2700 &  $-2.3\times 10^{-2}$\\ \hline
0.3  &  4000 &   800   &  800 &  $-1.7\times 10^{-2}$\\ \hline
0.4  &  2700 &   --    &  5300 & $2.0\times 10^{-2}$\\ \hline
\end{tabular}
\end{center}
\label{tab:T14}
\end{table}

For small $\Delta T$'s, $\Delta \chi''/\chi''$
is negative and $\chi''$ approaches the reference curve from below. 
As $\Delta T$ increases, this ratio
becomes positive meaning that the approach takes place now from above 
the reference curve (Fig. \ref{Fig:2}). For 
larger $\Delta T$'s, $\Delta \chi''/\chi''$ shows a maximum and decreases back towards zero. 
Beyond
that point, rejuvenation and full memory effects are observed 
\cite{MemChaos} and aging at
$T_2$ has no influence on the aging at $T_1$.
This characteristic oscillation of $\Delta \chi''/\chi''$ as a 
function of $\Delta T$ is the central result of this paper, 
that we discuss below in the context of Random Energy Models.
It is worth noticing that in the measurement of d.c. magnetization with $T$-cycle, 
the relaxation rate $S(t)$ at a short fixed time show a similar 
non monotonous behavior on $\Delta T$\cite{Nordblad}. This is quite reasonable 
if we notice the rough relation $\chi''(\omega)\sim S(\omega^{-1})$ 
(we thank to Nordblad for telling us this point).

The time $t^*$ at which the maximum of $\chi''$ occurs 
rapidly decreases as $\Delta T$ increases, whereas the recovery time $t_{\rm rec.}$ 
at which 
the signal merges with the shifted reference curve has a non monotonous
behavior with $\Delta T$ (see Tables 1 and 2). 
For still larger $\Delta T$'s, this time 
decays back to zero. In addition, quite surprisingly, $t_{\rm rec.}$ is very long 
for intermediate $\Delta T$'s -- much longer than expected from activated 
thermal slowing down. This {\it second non monotonous behavior} 
of $t_{\rm rec.}$ and the unexpected long $t_{\rm rec.}$ 
will be reconsidered in section \ref{sec:discussion}.

%\begin{table}
%\caption[]{\small $t_{\rm{eff}}$: shift time,  $t^*$: position of the maximum of 
%$\chi''$, $t_{\rm rec.}$: time at which the signal merges with the shifted 
%reference curve, and $\Delta \chi''/\chi''$: amplitude of the 
%memory anomaly for different $\Delta T$'s, for $T_1=14$ K. 
% }
%\begin{center}
%\begin{tabular}{||c|c|c|c|c||} \hline\hline
%$\Delta T$ (K) &  $t_{\rm{eff}}$ (sec) & $t^*$ (sec) & $t_{\rm rec.}$ (sec) &
%$\Delta \chi''/\chi''$ \\ \hline
%0.1 &  12000 &  1800  &  4200 &  $-2.4\times 10^{-2}$\\ \hline
%0.2  &  4000 &  900  &  900 &  $-2.6\times 10^{-2}$\\ \hline
%0.3  &  1500 &   160   &  8000 &  $2.8\times 10^{-2}$\\ \hline
%0.4  &  1000 &   --    &  10350 &  $9.1\times 10^{-2}$\\ \hline
%\end{tabular}
%\label{tab:T14}
%\end{center} 
%\end{table}

%\begin{table}
%\caption[]{\small Same caption as Table~1 for $T_1=12$ K.}
%\begin{center}
%\begin{tabular}{||c|c|c|c|c||} \hline\hline
%$\Delta T$ (K) & $t_{\rm{eff}}$ (sec) & $t^*$ (sec) &
%$t_{\rm rec.}$  (sec)
%&
%$\Delta \chi''/\chi''$ \\ \hline
%0.1 &  15000 &  1200  &  2850 &  $-1.8\times 10^{-2}$\\ \hline
%0.2  &  8000 &  1200  &  2700 &  $-2.3\times 10^{-2}$\\ \hline
%0.3  &  4000 &   800   &  800 &  $-1.7\times 10^{-2}$\\ \hline
%0.4  &  2700 &   --    &  5300 & $2.0\times 10^{-2}$\\ \hline
%\end{tabular}
%\end{center}
%\label{tab:T14}
%\end{table}

\section{Models}\label{sec:Model}
This section is devoted to introducing the REM, the GREM and their dynamical
extensions. A magnetization-like variable is introduced in order to define and
measure an ac-susceptibility. 
 
\subsection{The REM}
\begin{figure}
\centerline{\hbox{\epsfig{figure=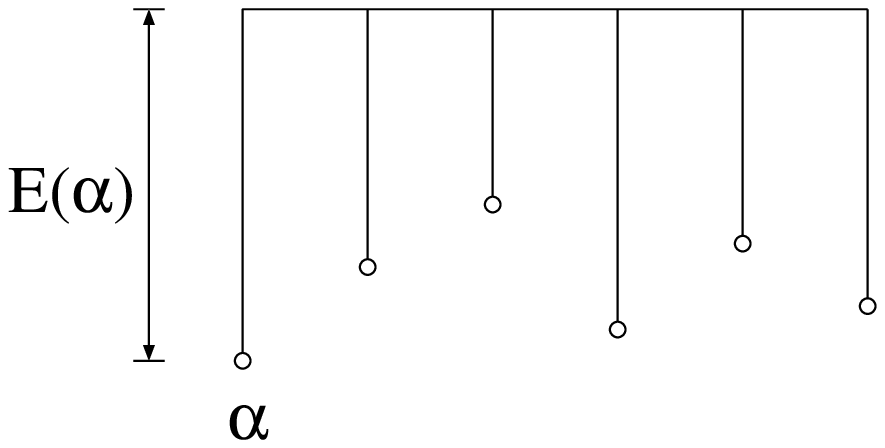,width=8.4cm}}}
\caption{Structure of the Random Energy Model.}
\label{fig:REM}
\end{figure}

The REM is schematically shown in Fig.~\ref{fig:REM}. 
The bottom points are the accessible states of the system. 
We will consider the case where $N$, the total number of states, 
is very large. Each branch represents the barrier energy \( E \), 
over which the system goes from one state to another. 
The values of $E$ are assigned randomly and independently 
according to the distribution :
\begin{equation}
\rho (E) {\rm d}E = \frac{{\rm d} E}{T_g}\exp[-E/T_g ]
\hspace{1cm}(E\ge 0),
\label{eqn:simplerho}
\end{equation}
where \( T_g \) is the transition temperature of the model. 
Hereafter \( T_g \) is used as the unit of temperature 
and is set to \( 1 \).

From the Arrhenius law, 
the escape time \( \tau(\alpha) \)
is related to \( E(\alpha) \) as
\begin{equation}
\tau({\alpha}) = \tau_0 \exp[ E(\alpha)/T],
\label{eqn:DefofTau}
\end{equation}
where $T$ is the temperature and \( \tau_0 \) is a microscopic time scale. 
Hereafter \( \tau_0 \) 
is used as the unit of time and is set to \( 1 \). From 
eq.~(\ref{eqn:simplerho}), the distribution of \( \tau \) is
\begin{equation}
p_x(\tau) {\rm d} \tau = \frac {x }{\tau^{x+1}}{\rm d} \tau 
\hspace{1cm}(\tau \geq 1),
\label{eqn:simpleptau}
\end{equation}
where \( x \equiv T/T_g \).
From eq.~(\ref{eqn:simpleptau}), 
it is clear that the averaged relaxation time 
is \( x/(x-1) \) for \( x>1 \) and infinite for \( x \le 1 \). 
This means that the transition from an ergodic phase to a non-ergodic phase 
occurs at \( T_g \)\cite{GREM1}. 

We define the dynamics of the REM from a simple Markoff process that defines a
`trap' model (see also \cite{BenArous}). At $t=0$, an initial state $\beta$ is chosen 
according to the uniform distribution over all states, i.e., 
\begin{equation}
P_{\alpha}(0)=\frac1N. 
\label{eqn:initialcondition}
\end{equation}
This means that the system is quenched from an 
infinitely high temperature. After the initial state is chosen, 
the system successively changes its state by repeating 
the following two processes. 
\begin{itemize}
\item[1.] The system is activated from the present state $\beta$ with 
 probability $\tau(\beta)^{-1}$ per unit time. 
\item[2.] After the activation from $\beta$, 
the system falls in one of all the states with uniform probability. 
\end{itemize}

When a magnetic field $H(t)$ is applied, 
the energy of a state $\beta$ is shifted by $-H(t)M_{\beta}$ and the activation 
energy  changes from $E(\beta)$ to
$E(\beta)+H(t)M_{\beta}$, where $M_\beta$ is the magnetization of a state 
$\beta$. This is the only effect 
of the magnetic field that we consider. 
The values of the magnetizations $M_\beta$ are assigned randomly and independently from 
a given distribution $D(M)$ with zero mean. 

\subsection{The GREM}

\begin{figure}
\centerline{\hbox{\epsfig{figure=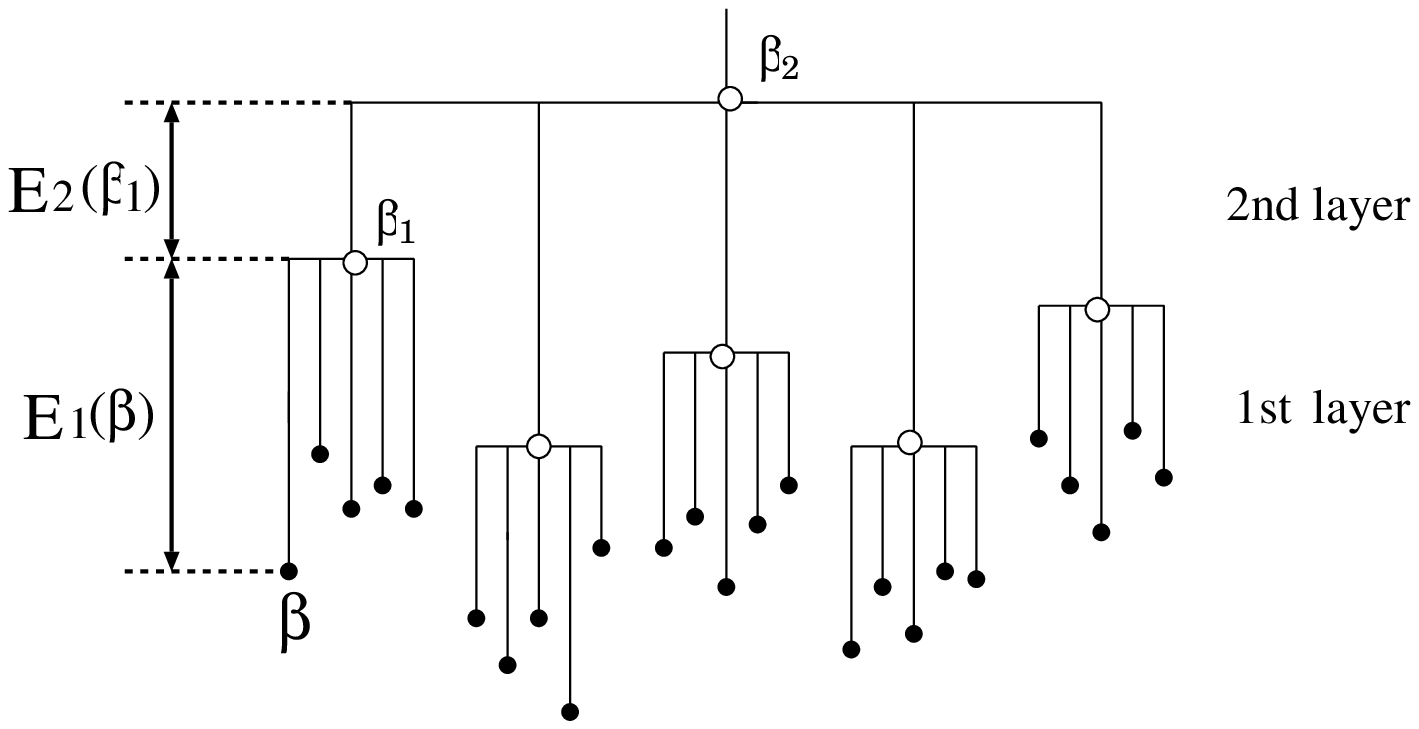,width=8.4cm}}}
\caption{Structure of the Generalized Random Energy Model with $L=2$. }
\label{fig:GREM}
\end{figure}
The GREM is schematically shown in Fig.~\ref{fig:GREM}. 
This model is generated by piling up $L$ different REM's in a hierarchical way. 
The Energy of a branch in the \( n \)-th layer 
(\( n \) is counted from the bottom), \( E_n \),
are given according to the distribution 
\begin{equation}
\rho_n(E_n){\rm d}E_n=\frac{{\rm d}E_n}{T_g(n)}
\exp[-E_n/ T_g(n)]
\hspace{5mm}(E_n\ge 0).
\end{equation}
The transition temperatures for the layers are chosen so as to satisfy
\begin{equation}
T_g(1)<T_g(2)<\cdots <T_g(L). 
\end{equation}
Therefore, in this model, the system freezes progressively from the
uppermost (the \( L \)-th) layer to the lowest one
as the temperature decreases. 

Now let us turn to the dynamics of the GREM. 
The initial state is given in the same way as the REM, i.e., 
eq.~(\ref{eqn:initialcondition}). 
After the initial state is chosen, 
the system successively changes its state by repeating 
the following two processes. 
\begin{itemize}
\item[1.] The system is activated from the present state $\beta$ 
to its $k$-th ancestor $\beta_k$ (see Fig.~\ref{fig:GREM}) with the probability
\begin{eqnarray}
W(\beta;k) &=&
 \left[\tau_0^{-1}\prod_{i=1}^k \exp[-E_i(\beta_{i-1})/T]\right] \nonumber \\
&&\times\biggl[1-\exp[-E_{k+1}(\beta_k)/T]\biggr],
\label{eqn:Wbetak}
\end{eqnarray}
per unit time. By convention, $E_{L+1}(\beta_L)\equiv\infty$. 
The first factor in the right hand represents 
the probability that the system is activated from \( \beta \) 
to \( \beta_k \) and the second one insures that the transition
from \( \beta_k \) to \( \beta_{k+1} \) is not active. 
\item[2.] After the activation from $\beta$ to $\beta_k$, 
the system falls to one of all the states ``under'' $\beta_k$ 
with uniform probability. 
\end{itemize}

When magnetic field $H(t)$ is applied, 
$E_1(\beta)$ in eq.~(\ref{eqn:Wbetak}) is replaced by 
$E_1(\beta)+H(t)M_{\beta}$. In order for nearby states 
to have strongly correlated magnetizations, 
the value of magnetization of a state $\beta$ 
is assigned to be
\begin{equation}
M_{\beta} ={\cal M}_1(\beta)+{\cal M}_2(\beta_1)+\cdots
+{\cal M}_{L}(\beta_{L-1}),
\end{equation}
where \( {\cal M}_{k+1}(\beta_k) \) is the contribution 
from the branching point $\beta_k$. The value of \( {\cal M}_{k} \) 
is assigned independently and randomly from a given distribution 
$D_k({\cal M}_k)$ with zero mean. If the distance $d(\alpha,\beta)$ 
between $\alpha$ and $\beta$ is $k$, 
i.e., $\alpha_n=\beta_n$ for $n\ge k$, 
the correlation between $M_\alpha$ and $M_\beta$ comes from the 
common contributions of ${\cal M}_n$ $(n\ge k+1)$ 
to these magnetizations, and is given by
\begin{equation}
\overline{M_{\alpha}M_{\beta}}=\sum_{n=k+1}^{L}
\overline{{\cal M}_n^2},
\end{equation}
where $\overline{{\cal M}_n^2}$ is the variance of $D_n({\cal M}_n)$. 
It decreases monotonically as $k$ increases and thus as 
the barrier between the two states 
becomes higher, just as occurs in the SK model\cite{Nemoto}. 

\section{Estimates of the a.c.-susceptibility}\label{sec:measurement}
Before we show our results, let us explain how we measure the 
ac-susceptibility. One simple way is to perform a Monte Carlo simulation. 
But time scales for which 
we can study by Monte Carlo simulation is quite different from 
the experimental one. For example, if we measure 
the period of the applied ac-field in units of 
the microscopic time of the system, 
a typical value in numerical studies is 
$10^2$ (see refs.~\cite{GREM3,GREM6,Yoshino:Takayama,Picco1,Picco2}), 
while that in experiments is $10^{6}-10^{12}$
(in the vicinity of $T_g$, the microscopic time scale may be renormalized 
by critical fluctuations: see refs.~\cite{SGactc1,SGactc2,SGactc4}). 

In the case of the REM, we can overcome this problem by using the relation
\begin{equation}
\chi(\omega,t)\approx \frac{\overline{M^2}}{T}\int 
\frac{1}{1-{\rm i}\omega \tau}Q(\tau,t){\rm d}\tau,
\label{eqn:relation}
\end{equation}
where $\omega$ is the angular frequency of the ac-field, 
$\overline{M^2}$ is the variance of $D(M)$ 
and $Q(\tau,t)$ is the probability 
density that the system is found at time \( t \) 
in one of the states whose relaxation time is \( \tau \). 
In eq.~(\ref{eqn:relation}), both $\chi(\omega,t)$ and 
$Q(\tau,t)$ are the disorder averaged quantities. 
The merit of this method is that we can calculate 
$Q(\tau,t)$ at arbitrary time $t$ for arbitrary initial condition $Q(\tau,t=0)$
because the Green function $G_{\beta\alpha}(t)$, i.e., 
the probability that the system which initially is 
at \( \alpha \) reaches \( \beta \) at time \(t\), 
has already been calculated analytically in ref.~\cite{GREM2}. 
As a result, we can estimate ac-susceptibility even for very long 
time scales comparable to those in experiments. 
The details of how we can calculate $Q(\tau,t)$ are described in the appendix. 

\begin{figure}
\centerline{\hbox{\epsfig{figure=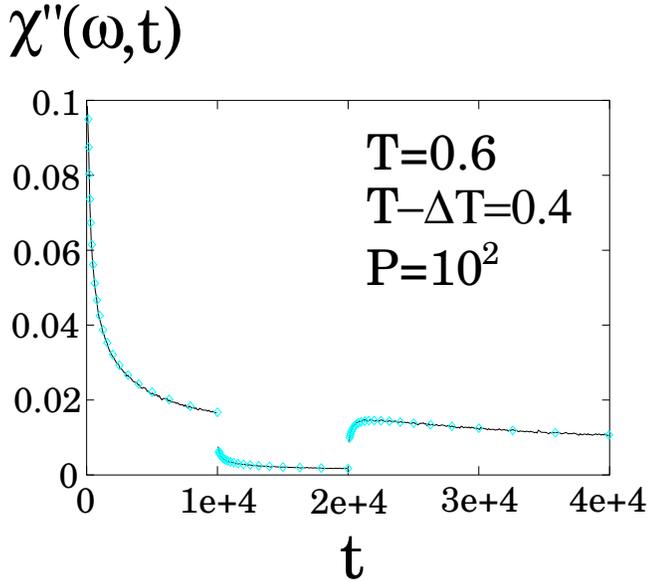,width=8.4cm}}}
\caption{The REM out-of-phase ac-susceptibility $\chi''(\omega,t)$ after 
	a negative $T$-cycle. It is measured in two different ways. 
	One can measure $\chi''(\omega,t)$ directly 
	by a Monte Carlo simulation (the lines). The another way is to measure
	$\chi''(\omega,t)$ using the relation 
	eq.~(\ref{eqn:relation}) (the diamonds). 
	After the system is quenched from an infinitely high temperature, 
	the temperature is changed as 
	$T_1=0.6\rightarrow T_2=0.4 \rightarrow T_1$. 
	The period of the applied ac-field $P$ is $100$. 
	For the data obtained by Monte Carlo simulation, 
	an average is taken over $2\times 10^7$ samples. 
}
\label{fig:check}
\end{figure}

Here the question is whether the relation 
eq.~(\ref{eqn:relation}) is valid or not 
even if the system is not equilibrated. 
In order to examine this question, 
we compared data obtained by Monte Carlo simulation 
and those obtained by eq.~(\ref{eqn:relation}). 
One example is shown in Fig.~\ref{fig:check}. 
A negative $T$-cycle is applied during the measurement. 
The agreement between both data is almost perfect. 
We also checked that both data coincide very well also for the 
in-phase ac-susceptibility $\chi'$. We can therefore trust the
validity of eq.~(\ref{eqn:relation}) for our purposes. 

Concerning the GREM, $\chi(\omega,t)$ has been measured by 
Monte Carlo simulation because we have not succeeded in 
calculating $Q(\tau,t)$ analytically. 

\begin{figure}
\centerline{\hbox{\epsfig{figure=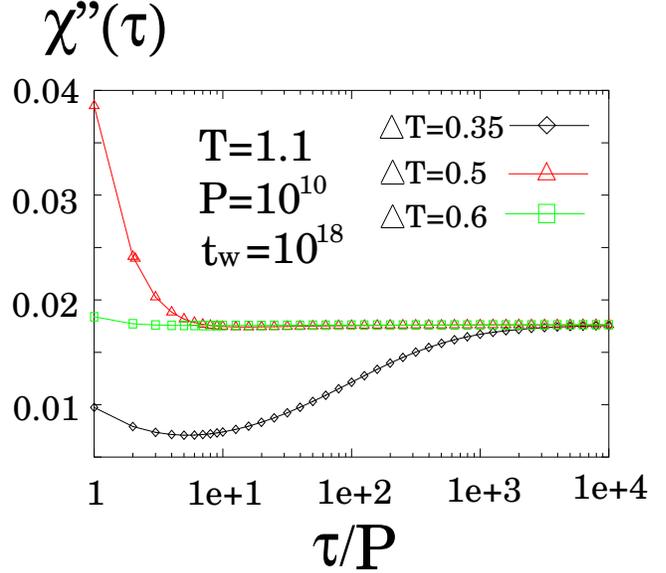,width=8.4cm}}}
\caption{The REM out-of-phase ac-susceptibility $\chi''$ after a negative 
        $T$-cycle, plotted 
	as a function of $\tau/P$, where 
	$\tau$ is the time elapsed after the temperature is returned to $T_1$ 
	and $P$ is the period of the applied ac-field. 
	After the system is quenched from an infinitely high temperature, 
	it is kept at $T_1=1.1$ for $t_{\rm w}=10^{18}$. 
	Then the temperature is reduced to $T_1-\Delta T$ 
	for $t_{\rm w}$, and is then shifted back to $T_1$. 
	The period of the applied ac-field is $P=10^{10}$. }
\label{fig:firstdata}
\end{figure}

\section{Results for the REM}\label{sec:resultREM}
In this section, the results of ac-susceptibility measurements during a $T$-cycle 
in the REM are shown. The measurement is done in the following way: 
After the system is quenched from an infinitely high temperature, 
the temperature is kept at $T_1$ in the first stage.
In the subsequent second stage the temperature is 
reduced to $T_1-\Delta T_n$ ($\Delta T_n=0.025 n$), and then it is 
returned to $T_1$ in the third stage. 
The time $t_{{\rm w}1}$ of the first stage and that of the second stage ($t_{{\rm w}2}$) 
are taken to 
be equal for simplicity: $t_{{\rm w}1}=t_{{\rm w}2}=t_{\rm w}$. 
The ac-susceptibility $\chi(\omega,t)$ is estimated by (\ref{eqn:relation}) 
with $\overline{M^2}=1$.

\subsection{The case $T_1 > T_g$}

\subsubsection{Results}

\begin{figure}
\centerline{\hbox{\epsfig{figure=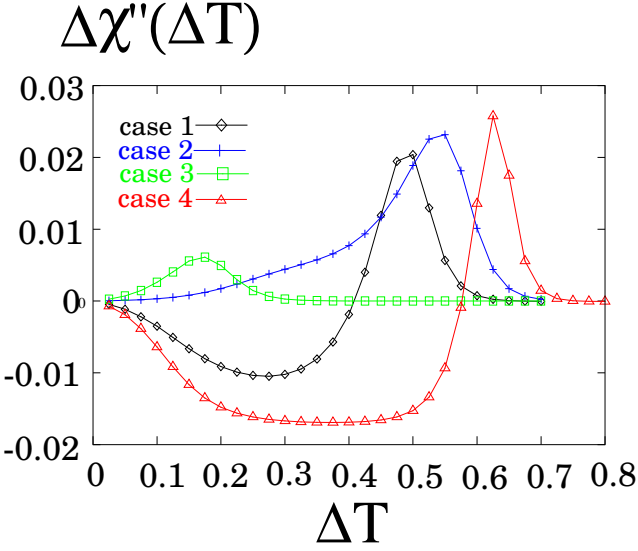,width=8.4cm}}}
\caption{$\Delta \chi''$ measured in the REM, plotted as a function of 
	$\Delta T$ (see eq.~(\ref{eqn:defofDelChi}) 
	for the definition of $\Delta \chi''$). 
	The values of the temperature $T_1$, the period of the applied ac-field $P$ 
	and the waiting time $t_{\rm w}$ are 
	$(T_1,P,t_{\rm w})=(1.1,10^{10},10^{18})$ for case 1, 
	$(T_1,P,t_{\rm w})=(1.25,10^{10},10^{18})$ for case 2, 
	$(T_1,P,t_{\rm w})=(1.1,10^{15},10^{18})$ for case 3 and 
	$(T_1,P,t_{\rm w})=(1.1,10^{10},10^{23})$ for case 4.
}
\label{fig:Pdependence}
\end{figure}

In Fig.~\ref{fig:firstdata}, 
out-of-phase ac-susceptibility $\chi''$ for $T_1=1.1$ and 
$t_{\rm w}=10^{18}$ is plotted as a function of $\tau/P$, 
where $\tau$ is elapsed time in the third stage and 
$P$ is the period of the applied ac-field. 
The value of $P$ is $10^{10}$. 
We find that $\chi''$ approaches its equilibrium value 
$\chi_{\rm eq}''$ from below 
for small $\Delta T$, it approaches from above for intermediate $\Delta T$, 
and $\chi''\approx\chi_{\rm eq}''$ 
from the beginning of the third stage for large $\Delta T$. 
This is exactly what is observed in experiments. 
In Fig.~\ref{fig:Pdependence}, we plot
\begin{equation}
\Delta \chi''\equiv \chi''(t=2t_{\rm w}+P)-\chi''(t=t_{\rm w}),
\label{eqn:defofDelChi}
\end{equation}
as a function of $\Delta T$ by the diamonds (case 1).
We can easily find the similarity between this curve 
and the experimental one shown in Fig.~\ref{Fig:3}. 

Now let us change one of the three parameters $T_1$, $P$ and $t_{\rm w}$ 
and see how these changes affect the result.  
First, $\Delta \chi''$ when only $T_1$ is changed 
from $1.1$ to $1.25$ is shown in Fig~\ref{fig:Pdependence} by the crosses 
(case 2). 
The behavior is quite different from that observed in the case 1 above, i.e., 
$\Delta \chi''> 0$ for all $\Delta T$'s. 
Next, $\Delta \chi''$ when only $P$ 
is changed from $10^{10}$ to $10^{15}$ 
is shown by the squares (case 3). 
We again find that $\Delta \chi'' > 0$ for all $\Delta T$'s. 
Finally, $\Delta \chi''$ when only $t_{\rm w}$ is changed 
from $10^{18}$ to $10^{23}$ is shown by the triangles (case 4). 
Although the value of $\Delta T$ above which $\Delta \chi'' \approx 0$ 
becomes large, there are not qualitative differences 
in comparison with case 1. Therefore, the non monotonous transient effect
disappears (i) at low frequencies and (ii) when the initial temperature is 
not close enough to $T_g$. 

\subsubsection{Qualitative discussion}

\begin{figure}
\centerline{\hbox{\epsfig{figure=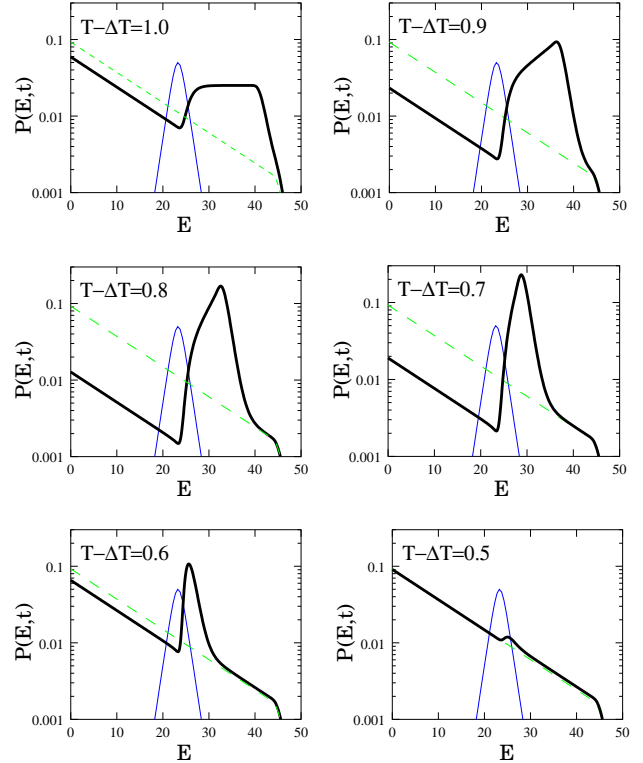,width=8.4cm}}}
\caption{The energy distribution $P(E,t)$ in the REM at time $t=2t_{\rm w}+P$ 
	after a negative $T$-cycle (the thick line). The values of 
	parameter $T$, $P$ and~$t_{\rm w}$ are the same as those of the case 1 
	in Fig.~\ref{fig:Pdependence}. 
	For comparison, a function proportional to 
	$\Omega (E)$ (see eq.~(\ref{eqn:DefofOmega})) and $P(E,t_{\rm w})$ 
	at $t=t_{\rm w}$ are drawn by the thin line and the broken one, 
        respectively. }
\label{fig:PEfig}
\end{figure}

In order to understand these surprising results,
let us investigate the time dependent energy distribution $P(E,t)$ 
which is related to $Q(\tau,t)$ through the relation
\begin{equation}
P(E,t) |{\rm d}E|=Q(\tau,t) |{\rm d}\tau|, 
\end{equation}
and eq.~(\ref{eqn:DefofTau}). 
In Fig.~\ref{fig:PEfig}, $P(E,t)$ at $t=2t_{\rm w}+P$ (i.e. in the
third part of the cycle, after one period of the a.c. field) is plotted 
for six different values of $\Delta T$ (the thick line).
The parameters $T_1$, $t_{\rm w}$ and $P$ 
are the same as those of the case 1 in Fig.~\ref{fig:Pdependence}. 
For comparison, a function which is proportional to 
\begin{equation}
\Omega(E)\equiv\frac{\omega \tau}{1+(\omega \tau)^2}=
\frac{(\frac{2\pi}{P})\exp(E/T)}{1+(\frac{2\pi}{P})^2\exp(2E/T)},
\label{eqn:DefofOmega}
\end{equation}
and $P(E,t)$ at $t=t_{\rm w}$ are drawn by the thin line and the broken one, 
respectively. It is worth noticing that the data of case 1 
in Fig.~\ref{fig:Pdependence} 
are obtained by integrating 
$P(E,t=2t_{\rm w}+P)\Omega(E)$ over $E$. From this figure, we find that 
$P(E,t)$ has both a minimum and a maximum as a function of $E$
(or a plateau in the special case $T_1-\Delta T=1.0$). 

When the system is kept at a temperature $T$ for a time $t$, 
the equilibration at $T$ proceeds and 
$P(E,t)$ becomes proportional to $\exp[\lambda_{\rm eq}(T) E]$ with 
\begin{equation}
\lambda_{\rm eq}(T)=\frac{1}{T}-\frac{1}{T_g},
\end{equation}
for $0\le E\le T\log(t)$. Therefore, $P(E,t)$ is equilibrated up to
the energy:
\begin{equation}
E_2 \approx (T_2)\log(t_{\rm w}),
\end{equation}
in the second stage. When the temperature is returned to $T_1$, 
the re-equilibration at temperature $T_1$ starts and it proceeds up to 
\begin{equation}
E_3 \approx T\log(P),
\end{equation}
at time $t=2t_{\rm w}+P$. These considerations naturally lead us 
to the approximate shape of the energy distribution:
\begin{equation}
P(E,2t_{\rm w}+P)\propto\left\{ 
  \begin{array}{cl}
\exp[\lambda_{\rm eq}(T_1) E]
&\mbox{($ 0\le E \le E_3 $)}, \vspace{2mm}\\
\exp[\lambda_{\rm eq}(T_2) E]
&\mbox{($ E_3\le E \le E_2 $)},
\end{array}\right.
\end{equation}
provided $E_3<E_2$. Accordingly, a minimum around $E\approx E_3$
and a maximum around $E \approx E_2$ appear for $T_2 < 1.0$, 
and there is a plateau between $E_3$ and $E_2$ for $T_2=1.0$. 
As for the case $T_2=0.5$, the peak is erased since in this case $E_2<E_3$. 

\begin{figure}
\centerline{\hbox{\epsfig{figure=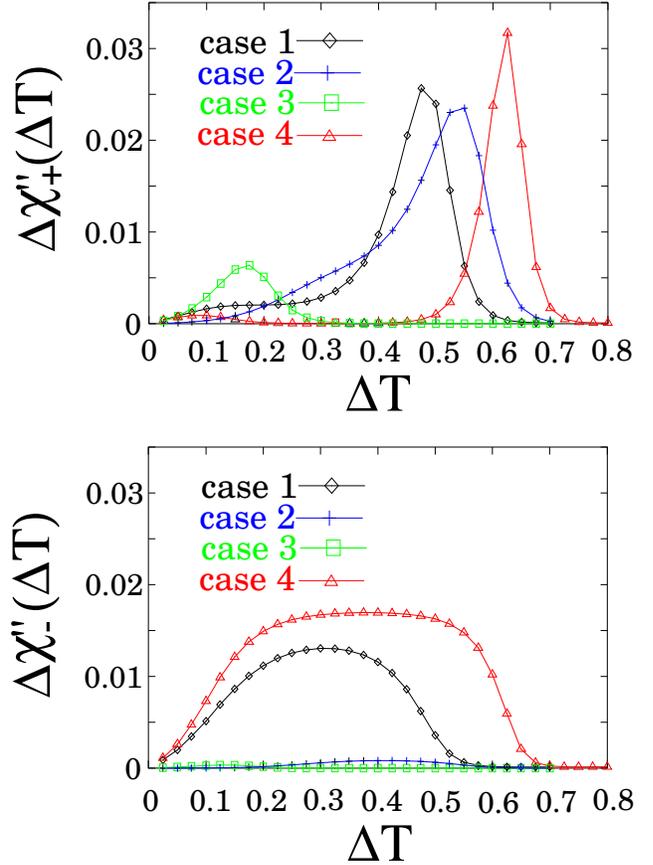,width=8.4cm}}}
\caption{The functions $\Delta\chi''_{+}(\Delta T)$ and 
	$\Delta\chi''_{-}(\Delta T)$ in the REM for each of the cases of 
	Fig.~\ref{fig:Pdependence} are plotted 
	as a function of $\Delta T$.
	See eqs.~(\ref{eqn:DefofDC-}) and (\ref{eqn:DefofDC+}) for 
	the definitions of $\Delta \chi_{+}''$ and 
	$\Delta \chi_{+}''$. }
\label{fig:separated}
\end{figure}

From Fig.~\ref{fig:PEfig}, the result of the case 1 
in Fig.~\ref{fig:Pdependence} 
is understood as follows. Because of the existence of a peak (or a plateau) 
and the normalization of $P(E,t)$ with respect to $E$, the difference 
$P(E,t=2t_{\rm w}+P)-P(E,t=t_{\rm w})$ changes its sign 
at a certain point $E_*$, which is slightly greater than $E_3$. 
Figure~\ref{fig:PEfig} shows that the peak around $E_2$ 
grows as $\Delta T$ increases. 
If we take the normalization condition into account, 
we notice that the growth of the peak means a decrease of 
$P(E,t=2t_{\rm w}+P)$ for $E<E_*$, and a corresponding increase of 
the quantity $\Delta\chi''_{-}(\Delta T)$, defined as
\begin{eqnarray}
\Delta\chi''_{-}(\Delta T)&\equiv& \int_0^{E_*} {\rm d}E \Omega(E)
\biggl|P(E,2t_{\rm w}+P) -P(E,t_{\rm w})\biggr|.\nonumber\\
\label{eqn:DefofDC-}
\end{eqnarray}
On the other hand, if $\Delta T$ is not very large and 
the location of the peak is not close to $E_3$, 
the second contribution $\Delta\chi''_{+}(\Delta T)$, defined as:
\begin{equation}
\Delta\chi''_{+}(\Delta T) \equiv \int_{E_*}^{\infty} {\rm d}E \Omega(E)
\biggl|P(E,2t_{\rm w}+P) -P(E,t_{\rm w})\biggr|,
\label{eqn:DefofDC+}
\end{equation}
cannot be very large (note that the peak of $\Omega(E)$ is around $E_3$). 
As a result, $\Delta \chi''=\Delta\chi''_{+}-\Delta\chi''_{-}$
is negative for small $\Delta T$ (see Fig.~\ref{fig:separated} 
where $\Delta\chi''_{+}(\Delta T)$ and $\Delta\chi''_{-}(\Delta T)$ 
for case 1 are plotted by the diamonds). 
Then, as $\Delta T$ increases, the location of 
the peak approaches $E_3$ and $\Delta\chi''_{+}(\Delta T)$ increases. 
Therefore, 
$\Delta \chi''(\Delta T)$ is positive for intermediate $\Delta T$. 
This peak disappears when $\Delta T$ is large and 
$E_2<E_3$. This is the reason why 
$\Delta \chi''(\Delta T)\approx 0$ for large $\Delta T$. 

Our next interest is to understand the following two trends:
\begin{itemize}
\item[(i)] If either the temperature or the period of the applied ac-field 
is large enough, $\Delta \chi''(\Delta T) > 0$ for all $\Delta T$. 
\item[(ii)] The behavior of $\Delta\chi''(\Delta T)$ does not 
depend sensitively on $t_{\rm w}$, 
except that the value of $\Delta T$ 
above which $\Delta \chi''\approx 0$ increases with $t_{\rm w}$. 
\end{itemize}
To understand these trends, an important point is to note that 
$\Delta\chi''_{-}(\Delta T)$ satisfies the inequality
\begin{eqnarray}
\Delta\chi''_{-}(\Delta T) &<& \int_0^{\infty} {\rm d}E \Omega(E)
P(E,t=t_{\rm w}) \nonumber \\
&\approx& \int_0^{\infty}{\rm d}E \lambda_{\rm eq}(T)
\exp[\lambda_{\rm eq}(T) E]\Omega(E)\nonumber \\
&\sim& P^{\frac{T_g-T}{T_g}}.
\label{eqn:inequality}
\end{eqnarray}
%for $T>T_g$. 
It is obvious from this inequality that 
$\Delta \chi''_{-}(\Delta T)$ is a decreasing function of 
$T$ and $P$. The quantity $\Delta\chi''_{+}(\Delta T)$, on the 
other hand, does not depend strongly on either $T$ or $P$. 
This explains point (i) above.

On the other hand, there is no $t_{\rm w}$ dependence 
in the inequality~(\ref{eqn:inequality}). 
This is the reason why behavior of $\Delta\chi''(\Delta T)$ 
does not depend on $t_{\rm w}$ so much. 
However, the value of $\Delta T$ 
above which $\Delta \chi''\approx 0$ 
increases with increasing $t_{\rm w}$ because 
it is determined from the condition:
\begin{equation}
E_2\approx E_3. 
\label{eqn:E2E3}
\end{equation}
The explanation for these two trends is confirmed by Fig.~\ref{fig:separated}, 
where $\Delta\chi''_{+}(\Delta T)$ and $\Delta\chi''_{-}(\Delta T)$ 
in each of the cases of Fig.~\ref{fig:Pdependence} are plotted 
as a function of $\Delta T$. The function $\Delta\chi''_{+}(\Delta T)$
does not depend much on the parameters $T$, $P$ and $t_{\rm w}$ 
compared to $\Delta\chi''_{-}(\Delta T)$, which becomes rather small 
when either $T$ or $P$ is large enough. 

\subsection{The case $T_1 < T_g$}

\begin{figure}
\centerline{\hbox{\epsfig{figure=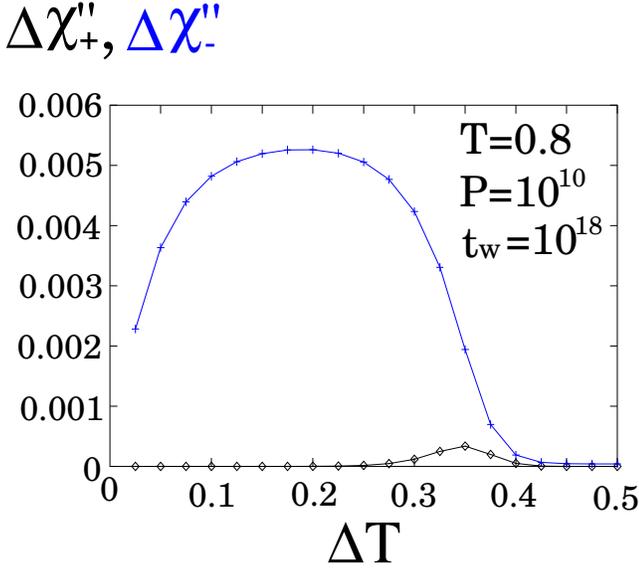,width=8.4cm}}}
\caption{The functions $\Delta \chi_{+}''$ (the diamonds) and 
	$\Delta \chi_{-}''$ (the crosses) in the REM for the case 
        $T_1=0.8$, $P=10^{10}$ and 
	$t_{\rm w}=10^{18}$. 
	The definition of $\Delta \chi_{+}''$ and that of $\Delta \chi_{-}''$ 
	are slightly changed by replacing $P(E,t=t_{\rm w})$ 
	in eqs.~(\ref{eqn:DefofDC-}) and~(\ref{eqn:DefofDC+}) with 
	$P_{\rm const}(E,t=t_{\rm eff})$, where $P_{\rm const}(E,t)$ is 
	isothermal energy distribution at $T$ and $t_{\rm eff}$ is defined 
	by eq.~(\ref{eqn:DeltaChiDef2}). }
\label{fig:belowTc}
\end{figure}

Figure~\ref{fig:belowTc} shows $\Delta \chi_{+}''$ and 
$\Delta \chi_{-}''$ in the case 
$T_1=0.8$, $P=10^{10}$ and $t_{\rm w}=10^{18}$. 
Because $\chi''$ decreases towards zero with time for $T<T_g$, 
we change slightly the definition of $\Delta \chi_{+}''$ and that of $\Delta \chi_{-}''$ 
and replace $P(E,t=t_{\rm w})$ 
in eqs.~(\ref{eqn:DefofDC-}) and~(\ref{eqn:DefofDC+}) with 
$P_{\rm const}(E,t=t_{\rm eff})$, where $P_{\rm const}(E,t)$ 
is isothermal energy distribution at $T_1$, and 
$t_{\rm eff}$ is estimated as 
\begin{equation}
t_{\rm eff}=t-t_{\rm w}+(t_{\rm w})^{T_2/T_1}.
\label{eqn:DeltaChiDef2}
\end{equation}
It is worth noticing that the effective time in the second stage 
is estimated as $(t_{\rm w})^{T_2/T_1}$. It has been shown 
in ref.~\cite{GREM4} that this way to estimate the effective time 
works well in the REM. The result is similar to that 
of the case $T_1 > T_g$ shown in Fig.~\ref{fig:Pdependence} in the sense 
that $\Delta \chi_{-}''$ has a wide plateau and $\Delta \chi_{+}''$ has a peak 
around the value of $\Delta T$ at which eq.~(\ref{eqn:E2E3}) 
is satisfied (in this case the value is about 0.356). The 
only difference is that $\Delta \chi_{+}''$ now never exceeds 
$\Delta \chi_{-}''$, so that $\Delta \chi''$ is negative 
for all $\Delta T$. 

The dependence on the different parameters was also investigated by changing 
one of the three parameters $T$, $P$ and $t_{\rm w}$. 
The ranges we examined are 
$0.75\le T_1\le 0.9$, $10^{10}\le P \le 10^{15}$ and 
$10^{18}\le t_{\rm w} \le 10^{23}$, respectively. 
As a result, we found that $\Delta \chi_{-}''>\Delta \chi_{+}''$ is 
always satisfied for all $\Delta T$ in all the cases. 
Therefore, the condition $T_1 > T_g > T_2$ is required to observe a non
monotonous memory anomaly.

\section{Results for the GREM}\label{sec:resultGREM}

Now let us turn our attention to the GREM. 
As mentioned in \S \ref{sec:measurement}, 
$\chi(\omega,t)$ for the GREM has been measured using 
Monte Carlo simulation because we have not succeeded in 
calculating $Q(\tau,t)$ analytically. However, we do not 
measure $\chi(\omega,t)$ from the linear response to an ac-field because this 
procedure requires averaging over a very large number of samples 
(typically $10^7 - 10^8$ samples). Instead, we have estimated 
the ac-susceptibility from the relations:
\begin{equation}
\chi_k(\omega,t)=\frac{\overline{{\cal M}_k^2}}{T}\int 
\frac{1}{1-{\rm i}\omega \tau_k} Q_k(\tau_k,t){\rm d}\tau_k,
\label{eqn:relation2}
\end{equation}
where 
\begin{equation}
\tau_k\equiv \exp\left[\frac{\sum_{i=1}^k E_i}{T}\right],
\end{equation}
the function $Q_k(\tau,t)$ is the probability density of $\tau_k$ at time t 
and $\chi_k(\omega,t)$ is ac-susceptibility calculated from 
${\cal M}_k$. It is this probability density $Q_k(\tau_k,t)$ that we 
obtained from Monte Carlo simulation. 
The validity of these relations was confirmed numerically 
by comparing data from the direct measurement of ${\cal M}_k$ under an ac-field
and from eq.~(\ref{eqn:relation2}). 

\begin{figure}
\centerline{\hbox{\epsfig{figure=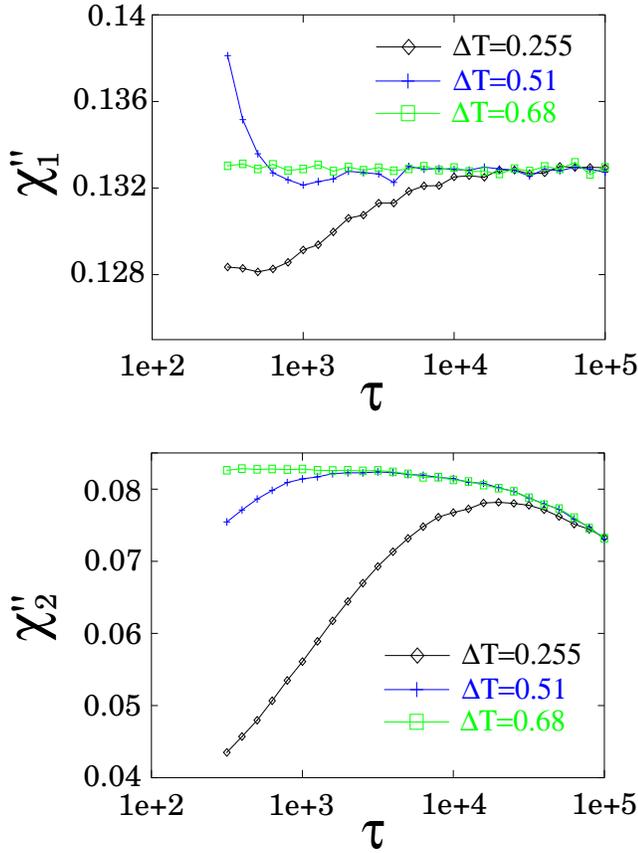,width=8.4cm}}}
\caption{The GREM out-of-phase ac-susceptibility $\chi_k''$ after a negative 
        $T$-cycle as a function of the elapsed time $\tau$ after 
	the temperature is returned to $T_1$.  This GREM has $L=2$, 
	$T_g(1)=0.5$ and $T_g(1)=1.0$. 
	The period of the applied ac-field 
	is $P=3\times 10^2$. After the system is quenched 
	from an infinitely high temperature, 
	the temperature is kept at $T_1=0.85$ for $t_{\rm w}=10^5$ 
	in the first stage. In the second stage 
	the temperature is temporally reduced to $T_1-\Delta T$ for the same 
	time $t_{\rm w}$, and before being  returned to $T_1$ in the last stage. 
}
\label{fig:GREMdata}
\end{figure}

Because we have to rely on Monte Carlo simulation, 
the time scales are rather restricted 
as compared to the REM. Therefore, we will confine 
ourselves to showing results with one set of parameters. 
The system we have investigated is the GREM with $L=2$ 
($L$ is the number of layers), $T_g(1)=0.5$ and 
$T_g(2)=1.0$. The disorder average is taken over $8\times 10^6$ samples. 
The period of the applied ac-field is $300$. 
After the system is quenched from an infinitely high temperature, 
the temperature is kept at $T_1=0.85$ for $t_{\rm w}=10^5$ in the first stage.
In the subsequent second stage the temperature is 
reduced to $T_2$ for $t_{\rm w}$, and then it is 
returned to $T_1$ in the third stage. In Fig.~\ref{fig:GREMdata}, 
the contribution from both levels, $\chi_1''$ and 
$\chi_2''$ are plotted as a function of $\tau$, where 
$\tau$ is elapsed time in the third stage. As for $\chi_1''$, 
we again find a non monotonous behavior, similar to that observed in 
experiments (Fig.~\ref{Fig:3}) and in the REM (Fig.~\ref{fig:firstdata}).
This was expected, since for the first level dynamics is very similar to 
the single REM, with transitions to the higher level frozen by the fact that
$T_1 < T_g(2)$.

On the other hand, the memory anomaly in  
$\chi_2''$ is always negative, for all values of $\Delta T$. 
This result is consistent with that obtained in the REM 
where $\chi''$ always approaches the reference curve from below if $T_1<T_g$ 
(note that $T_1<T_g(2)$ in the present case). 

Finally, let us discuss what would happen in the case $L\gg 1$. 
There, some layers are frozen 
($T<T_g(n)$) and others layers are {\it fast} ($T>T_g(n')$)
at any given temperature $T < T_g=T_g(L)$. 
From the study on the REM shown in \S \ref{sec:resultREM}, 
we expect that the contributions to $\chi''$ from the 
frozen layers will always lead to a negative memory anomaly,
where as the contribution from the `critical' levels 
will lead to a non monotonous contribution.

\section{Discussion. Other scenarios}\label{sec:discussion}

We have seen that the non monotonous transient effect observed in memory experiments can be
reproduced within simple REM trap model, provided the temperature is above and close enough to
the critical temperature and the frequency not too low. The same mechanism is present in
the GREM, and is governed by the dynamics around the `critical level', i.e. the level 
such that its critical temperature is close to the working temperature. As emphasized 
in refs.~\cite{StAndrews,PRB}, the physical interpretation of the different `levels' is in terms
of length scales: small scale dynamics corresponds to the deepest level of the tree, whereas
large length scales correspond to the upper level of the trees. The observed aging dynamics 
always concerns those length scales (levels) around the critical temperature: larger length 
scales are frozen, whereas smaller length scales are completely equilibrated. Hence, in the
above GREM interpretation, the important ingredient is that the system remains close to criticality 
at any temperature, but the basic ingredient is already present in the REM, and is related to 
the abrupt change of the way the different states are explored at $T_g$ (see \cite{marta}).

The GREM model is a concrete implementation of the so called `hierarchical' interpretation
of experimental data \cite{refregier}, to which one often opposes the `droplet' interpretation \cite{drop}. As
discussed in details in ref.~\cite{PRB}, the two interpretations are to some extent complementary if
one wants to interpret the `hierarchy' of phase space as a hierarchy of 
length and time scales.

However, the droplet interpretation of the rejuvenation and memory experiments 
makes an
extra assumption that we now discuss. 
The existence of an overlap length $\ell_{\Delta T}$ between typical 
configurations at $T$ and $T-\Delta T$ is postulated \cite{drop,chaosbm}, such that for length 
scales larger than $\ell_{\Delta T}$, the configurations 
at the two temperatures are completely 
unrelated (`temperature chaos'). Using plausible arguments, one 
deduces that $\ell_{\Delta T}$ should diverge as a
power-law of $\Delta T$ for $\Delta T \to 0$. Recent experimental data has 
given some credit to the existence of temperature chaos \cite{Yoshino:Suedes2}. 
After a waiting time $t_{{\rm w}1}$, the active length scales are such 
that $\tau_r(\ell_{{\rm w}1},T_1) \sim t_{{\rm w}1}$, where 
$\tau_r(\ell,T)$ is the typical relaxation time corresponding 
to length $\ell$ at temperature $T$. Length scales 
much smaller than $\ell_{{\rm w}1}$ are fully equilibrated. In this picture, the scenario for rejuvenation is thus the following: whenever $\ell_{{\rm w}1} < \ell_{\Delta T}$, the temperature
change does not modify the achieved pattern, but only acts to slow down the 
dynamics. Conversely, 
when $\Delta T$ is such that $\ell_{{\rm w}1} > \ell_{\Delta T}$, the system has to start rebuilding new correlations as if it were brought directly 
from high temperature (when 
$\ell_{\Delta T} \to 0$). 
As shown in details in ref.~\cite{Yoshino:Lemaitre}, this does not necessarily mean
 that the structure 
grown at the first temperature is immediately washed away. On the contrary, as long as the
length scales $\ell_{{\rm w}2}$, active at $T_2$, remain small compared to $\ell_{{\rm w}1}$, 
memory can be partially or totally recovered. 
The criterion is the following: the time needed
to erase the effect of the dynamics at $T_2$ 
when the system is heated back, and
isothermal dynamics at $T_1$ is recovered, is given by:
\begin{equation}\label{terase}
t_{\rm rec.}(t_{{\rm w}2}) \sim \tau_r(\ell_{{\rm w}2},T_1).
\end{equation}
Since $\ell_{{\rm w}2}$ decreases extremely fast with decreasing 
temperature \cite{PRB,SGactc4,BerthierJPB,Yoshino:Suedes,Yoshino:4D}, 
$t_{\rm rec.}(t_{{\rm w}2})$ 
decreases very rapidly (for a given $t_{{\rm w}2}$) as $\Delta T$ increases, and should soon become smaller than $\omega^{-1}$, which is the 
smallest time for which a measurement of
the a.c. susceptibility can be performed. 

When $t_{\rm rec.}(t_{{\rm w}2}) > \omega^{-1}$ and $\ell_{w2} > \ell_{\Delta T}$,
on the other hand, one expects to see an initial spike in 
the a.c. susceptibility that 
corresponds to the reconstruction of small length scale correlations at $T_1$. 
Schematically, the
temperature chaos scenario therefore predicts that the memory anomaly $\Delta \chi$ should be
zero for $\Delta T < \Delta T^*$, with $\ell_{{\rm w}1} = \ell_{\Delta T^*}$, positive 
for larger $\Delta T$, but becoming zero again when $t_{\rm rec.}(t_{{\rm w}2})$ becomes 
shorter than $\omega^{-1}$. 

One can finally argue that the number of thermally active 
(equilibrium) droplets decreases slightly when the temperature is reduced from $T_1$ to $T_2$,
thereby reducing the equilibrium a.c. susceptibility. The need to re-nucleate these droplets
back at $T_1$, which also takes a time $\sim t_{\rm rec.}$, would then explain the negative contribution to the memory anomaly for small $\Delta T$. \footnote{Note that the negative initial contribution to $\Delta \chi$ is like 
the Kovacs effect in polymeric glasses \cite{Kovacs,Yoshino:Takayama,BerthierJPB}.}
 This would predict that $\Delta \chi''/\chi''_{\rm{eq}} \simeq - [1/T
+ |\Upsilon'/\Upsilon|] \Delta T$ for
small $\Delta T$, where $\Upsilon$ is the temperature dependent stiffness of the droplets. 
The experimental effect, found to be stronger than $-\Delta T/T$, is in qualitative agreement with this prediction (see Fig.~\ref{Fig:3}). We have furthermore 
checked that the amplitude of the bump in 
$\chi''$ at $t=t_{\rm w1}+t_{\rm w2}$ is of the same order as the difference between the equilibrium values of 
$\chi''_{\rm{eq}}$ at $T_1$ and $T_1-\Delta T$.

However, the time $t_{\rm rec.}$ beyond which the stay at $T_2$ is erased does not
conform to the naive estimate eq.~(\ref{terase}), since it is found to be non monotonous and much larger than expected. It is rather the position $t^*$ of the maximum of $\chi''$ that seems to obey Eq. (\ref{terase}). 
Note that the REM scenario also predicts a monotonously decreasing recovery time $t_{\rm rec.}$ with
increasing $\Delta T$. 
We have at present no physical interpretation for this discrepancy.

The experimental data appears to be consistent both with the above droplet/chaos 
interpretation and with the hierarchical model developed in the present paper 
(see also the discussion in \cite{PRB,Yoshino:Suedes2}).
The present study shows explicitly that the non monotonous 
memory anomaly does not require the existence of an overlap length. 
Indeed, we argued that the 
REM trap model, where this overlap length is absent, is also able to 
reproduce 
qualitatively the memory anomaly if one works 
around the freezing temperature around 
which `temperature chaos' effects are observed \cite{marta}. 
In the REM scenario, the positive 
contribution to the memory anomaly comes from an over 
concentration of the probability weight in deep traps 
at $T_2$ as compared to the equilibrium 
situation at $T_1$ (see the discussion in ref.~\cite{marta}). 
Physically, this positive contribution
corresponds to a freezing at $T_2$ of small length scales that have to 
unfreeze when back at $T_1$, a scenario 
that was directly confirmed by the numerical simulations 
of \cite{BerthierJPB} where 
temperature chaos is absent but rejuvenation and memory effects 
are clearly observed. 

\vspace{1cm}

\centerline{{\bf Acknowledgments}}
We wish to thank L. Berthier, J. Hammann, 
O. Martin and H. Yoshino for many interesting conversations. We thank P. Nordblad
and H. Yoshino for very useful comments on the manuscript.
M. S. acknowledges a fellowship from the French Ministry of research. 
M. S also would like to thank CEA Saclay for hospitality where 
main part of this work was performed. The LPTMS laboratory is an Unit\'e de 
Recherche de l'Universit\'e Paris~XI associ\'ee au CNRS.

\appendix
\vspace{1cm}
\centerline{{\bf Appendix}}

\bigskip

This appendix is devoted to explain in detail how 
we can calculate $Q(\tau,t)$. 
We assume that the probability $P_{\alpha}(t=0)$ that the system 
is found at a state $\alpha$ at time $t=0$ is given. 
For simplicity, let us first consider the case that the system 
is kept at a constant temperature $T$. It is easily found that 
$Q(\tau,t)$ is given as 
\begin{equation}
Q(\tau,t){\rm d}\tau=\sum_{\beta,\alpha}{\rm d}\tau 
\delta(\tau({\beta}) -\tau) 
G_{\beta\alpha}(t)P_{\alpha}(t=0),
\label{eqn:DefofQtaut}
\end{equation}
where $G_{\beta\alpha}(t)$ is the Green function, 
i.e., the probability that the system which initially is 
at \( \alpha \) reaches \( \beta \) at time \(t\). 

Now let us calculate the Green function. 
When the system which initially is at \( \alpha \) reaches 
\( \beta \) at time \(t\), 
there are the following two possibilities:
\begin{itemize}
\item[(i)] \( \alpha=\beta \) and the system has not been activated 
during time \( t \). 
\item[(ii)] The system is activated at \( t' \) \( (<t) \) and 
reaches \( \beta \) after that time. 
\end{itemize}
In the case (ii), because the new state after the activation 
is chosen randomly from all the states, 
the probability that the system reaches \( \beta \) after the activation is 
$P_{\beta}^{\rm uni}(t-t')$, where
\begin{equation}
P_{\beta}^{\rm uni}(t)=\frac{1}{N}\sum_{\gamma} G_{\beta\gamma}(t),
\label{eqn:DefofPuni}
\end{equation}
and $N$ is the number of states. Taking this fact into consideration 
and recalling that the system is activated from $\alpha$ 
with the probability $\tau({\alpha})^{-1}$, we obtain
\begin{eqnarray}
G_{\beta\alpha}(t)
&=&\delta_{\alpha \beta}\exp\left[-\frac{t}{\tau(\alpha)}\right]\nonumber\\
&&+\int_0^{t}\frac{{\rm d}t'}{\tau(\alpha)} 
\exp\left[-\frac{t'}{\tau(\alpha)}\right]P_{\beta}^{\rm uni}(t-t').
\label{eqn:Green}
\end{eqnarray}
The Laplace transformation of this equation leads us to 
\begin{eqnarray}
{\hat G}_{\beta\alpha}(s) &\equiv& \int_0^{\infty}{\rm d}t \exp[-st]
G_{\beta\alpha}(t)\nonumber \\
&=& \frac{\tau(\alpha) \delta_{\alpha\beta} }{s\tau(\alpha)+1}
+\frac{{\hat P}_{\beta}^{\rm uni}(s)}{s\tau(\alpha)+1},
\end{eqnarray}
where ${\hat P}_{\beta}^{\rm uni}(s)$ is the Laplace transformation of 
$P_{\beta}^{\rm uni}(t)$. From this equation and 
the Laplace transformation of eq.~(\ref{eqn:DefofPuni}), 
we find
\begin{equation}
{\hat P}_{\beta}^{\rm uni}(s)=\frac{\displaystyle \frac{\tau(\beta)}{s\tau(\beta)+1}}
{\displaystyle \sum_{\alpha}\frac{s\tau(\alpha)}{s\tau(\alpha)+1}}.
\end{equation}
The calculation of ${\hat P}_{\beta}^{\rm uni}(s)$ for small $s$ 
and its inverse Laplace transformation have already been done 
in ref.~\cite{GREM2}. The results are
\begin{equation}
{\hat P}_{\beta}^{\rm uni}(s) = \left\{ 
  \begin{array}{cl}
\displaystyle{\frac{\tau(\beta)}{N x s^x c(x) (s\tau(\beta)+1) }}
&\mbox{($ x<1 $)}, \vspace{2mm}\\
\displaystyle{\frac{-\tau(\beta)}{N s \log(s) (s\tau(\beta)+1) }}
&\mbox{($ x=1 $)},\vspace{2mm}\\
\displaystyle{\frac{(x-1)(s+1)\tau(\beta)}{N x s (s\tau(\beta)+1) }}
&\mbox{($ x>1 $)},
\end{array}\right.
\vspace{5mm}
\end{equation}
and
\begin{equation}
P_{\beta}^{\rm uni}(t) = \left\{ 
  \begin{array}{cl}
\displaystyle{\frac{\int_0^t {\rm d} u u^{x-1} \exp[-(t-u)/\tau(\beta)]}
{N x c(x) \Gamma(x)}} &\mbox{($ x<1 $)}, \vspace{2mm}\\
\displaystyle{\frac{\tau(\beta)[1-\exp[-t/\tau(\beta)]]}{\log(t)}}
&\mbox{($ x=1 $)},\vspace{2mm}\\
\displaystyle{\frac{(x-1)\tau(\beta)\{1-\exp[-t/\tau(\beta)]\}}{N x }}
&\mbox{($ x>1 $)},
\end{array}
\right.
\label{eqn:resultPuni}
\vspace{5mm}
\end{equation}
where $x\equiv T/T_g$ and 
\begin{equation}
c(x)={\rm \Gamma}(x){\rm \Gamma}(1-x)=\frac{\pi}{\sin(\pi x)}.
\end{equation}

Now let us return to the calculation of $Q(\tau,t)$. The substitution of 
eq.~(\ref{eqn:Green}) into eq.~(\ref{eqn:DefofQtaut}) leads us to
\begin{eqnarray}
\hspace*{-7mm}Q(\tau,t)&=&\exp[-t/\tau]Q(\tau,0)\nonumber \\
\hspace*{-7mm}&&+\int_{1}^{\infty} {\rm d}\tau' \int_{0}^{t} \frac{{\rm d}t'}{\tau'}
Q^{\rm uni}(\tau,t-t'){\rm e}^{-\frac{t'}{\tau'}}
Q(\tau',0),
\label{eqn:Qrepresentation}
\end{eqnarray}
where
\begin{equation}
Q(\tau,0){\rm d}\tau\equiv \sum_{\alpha} {\rm d}\tau 
\delta(\tau(\alpha) -\tau) P_{\alpha}(t=0),
\end{equation}
and
\begin{eqnarray}
Q^{\rm uni}(\tau,t){\rm d}\tau &\equiv& \sum_{\alpha} {\rm d}\tau 
\delta(\tau(\alpha) -\tau) P_{\alpha}^{\rm uni}(t)\nonumber \\
&=& {\rm d}\tau N p_x(\tau) P^{\rm uni}(\tau,t).
\label{eqn:DefofQuni}
\end{eqnarray}
The function $p_x(\tau)$ is defined by eq.~(\ref{eqn:simpleptau}). 
From eqs.~(\ref{eqn:resultPuni}), (\ref{eqn:Qrepresentation}) and 
(\ref{eqn:DefofQuni}), we finally obtain
\begin{eqnarray}
&&Q(\tau,t)-\exp[-t/\tau]Q(\tau,0)\nonumber \\
&& = \left\{ 
  \begin{array}{cl}
\displaystyle{
\hspace*{-3mm}\frac{p_x(\tau)\tau}{x c(x) \Gamma(x) }
\int_{1}^{\infty} {\rm d}\tau' \int_0^{t} {\rm d} u
u^{x-1}\frac{Q(\tau',0)}{\tau'-\tau}}\nonumber \vspace{2mm}\\
\displaystyle{
\hspace*{-3mm}\times\biggl[\exp\left(\frac{u-t}{\tau'}\right)
-\exp\left(\frac{u-t}{\tau}\right)\biggr]} 
&\mbox{($ x<1 $)}, \vspace{2mm}\\
\hspace*{-27mm}\displaystyle{\frac{p_x(\tau)\tau}{\log(t)}\int_0^{\infty} 
{\rm d}\tau' Q(\tau',0)}\nonumber\vspace{2mm}\\
\displaystyle{
\hspace*{6mm}\times\Bigl[\bigl\{ 1-{\rm e}^{-\frac{t}{\tau'}} \bigr\}-\frac{\tau}{\tau'-\tau}
\bigl\{ {\rm e}^{-\frac{t}{\tau'}} -{\rm e}^{-\frac{t}{\tau}}\bigr\}\Bigr]}
%[ \{ 1-\exp(-t/\tau') \} -\frac{\tau}{\tau'-\tau}
%\{ \exp(-t/\tau') -\exp(-t/\tau) \}]}
&\mbox{($ x=1 $)},\vspace{2mm}\\
\hspace*{-17mm}\displaystyle{\frac{(x-1)p_x(\tau)\tau}{x}\int_0^{\infty} 
{\rm d}\tau' Q(\tau',0)}\nonumber \vspace{2mm}\\
\displaystyle{
\hspace*{6mm}\times\Bigl[\bigl\{ 1-{\rm e}^{-\frac{t}{\tau'}} \bigr\}-\frac{\tau}{\tau'-\tau}
\bigl\{ {\rm e}^{-\frac{t}{\tau'}} -{\rm e}^{-\frac{t}{\tau}}\bigr\}\Bigr]}
&\mbox{($ x>1 $)}.
\end{array}\right.\\
\vspace{5mm}
\label{eqn:finalformula}
\end{eqnarray}

Next, let us consider how we can calculate $Q(\tau,t)$ 
when the temperature is changed discontinuously as
\begin{equation}
T(t)=T_i\hspace{2cm}(t_{i}\le t \le t_{i+1}).
\end{equation}
The answer is rather simple. 
At first, we calculate $Q(\tau,t_{{\rm w}1})$ with some 
initial distribution $Q(\tau,0)$. Then, 
we set the new initial distribution to $Q(\tau,t_{{\rm w}1})$ and 
use eq.~(\ref{eqn:finalformula}) to calculate $Q(\tau,t_{{\rm w}2})$. 
We can calculate $Q(\tau,t)$ at any $t$ by repeating this procedure.

%%%%%%%%%%%%%%%%%%%%
\end{document}